# Theory of Coupled Resonator Optical Waveguides (CROW) Exhibiting High Order Exceptional Points of Degeneracy

Mohamed Y. Nada, Mohamed A. K. Othman, and Filippo Capolino

*Department of Electrical Engineering and Computer Science, University of California, Irvine, CA 92697, USA*

We present a novel approach and a theoretical framework for generating high order exceptional points of degeneracy (EPD) in photonic structures based on periodic coupled resonators optical waveguides (CROWs). Such EPDs involve the coalescence of Floquet-Bloch eigenwaves in CROWs, without the presence of gain and loss, which is in contrast to the requirement of Parity-Time (PT) symmetry to develop exceptional points based on gain and loss balance. The EPDs arise here by introducing symmetry breaking in a conventional chain of coupled resonators through coupling the chain of resonators to an adjacent uniform optical waveguide, which leads to unique modal characteristics that cannot be realized in conventional CROWs. Such remarkable characteristics include high quality factors (*Q*-factor) and strong field enhancement, even without any mirrors at the two ends of a cavity. We show for the first time the capability of CROWs to exhibit EPDs of various order; including the degenerate band edge (DBE) and the stationary inflection point (SIP). The proposed CROW of finite length shows enhanced quality factor when operating near the DBE, and the *Q*-factor exhibits an anomalous scaling with the CROW's length. We develop the theory of EPDs in such unconventional CROW using coupled-wave equations, and we derive an analytical expression for the dispersion relation. The proposed unconventional CROW concepts have various potential applications including *Q*-switching, nonlinear devices, lasers, and extremely sensitive sensors.

## I. INTRODUCTION

Confinement of light in optical microresonators (or microcavities) is one of the fundamental processes for enhancing optical interactions for various applications [1,2], including filters [3], sensors [4], optical delay line devices [5], optical switching and modulators [6,7], optical buffers [8], lasers [9], energy harvesting applications, and in engineering of the wave-matter interaction in quantum systems [10]. Photonic crystal cavities [11], disk [9], toroid [12], or ring [13] microcavities, among others, are typically employed in those aforementioned applications requiring high quality (*Q*)factor. On the other hand, cascading a chain of coupled micro resonators, as was introduced in [14], has stimulated a great interest in studying coupled resonator optical waveguides (CROWs) as efficient devices for light transport [5,15,16].

Slow-light phenomenon, whereby the group velocity of light in optical structures is low (much lower than the velocity of light in free space *c*) [17,18] has spawned many intriguing new aspects of optical resonators in which nonlinearities (harmonic generation, wave mixing, etc.) [19], and gain/absorption [20] among other features can be significantly enhanced. A particular kind of slow wave resonance occurs in the vicinity of the transmission band edge of periodic structures. For this reason, a slow wave resonance is often referred to as a transmission band edge resonance [21–23]. At the band edge, degeneracy of Floquet-Bloch eigenwaves that coalesce (in both eigenvalues and eigenvectors) at a single frequency. Degenerate band edge (DBE) [21,24,25] arises when four Floquet-Bloch eigenwaves coalesce in periodic structures supporting multiple polarization eigenwaves that are periodically mixed. The DBE condition, which is a fourth order eigenwave degeneracy, causes a quartic power dependence at the band edge of the dispersion diagram, $(\omega - \omega_d) \propto (k - k_d)^4$ where $\omega_d$ is the DBE angular frequency, $k$ is the Floquet-Bloch wavenumber, $k_d$ is the band edge wavenumber. Hence the DBE condition is accompanied by a significant reduction in the group velocity of waves and improvement in the in the local density of states. DBE has stimulated research in many interesting applications such as small antennas [26], low-threshold lasing and optical switching [25,27], and efficient high power microwave generation [28,29].

Furthermore, another important point of degeneracy is the stationary inflection point (SIP), which is a third order degeneracy [30] and in its proximity the dispersion relation follows $(\omega - \omega_s) \propto (k - k_s)^3$ where $\omega_s$ is the SIP angular frequency, $k$ is the Floquet-Bloch wavenumber, $k_s$ is the SIP Floquet-Bloch wavenumber. Slow light associated with the SIP shows promising characteristics [31] potentially useful for many applications. In general, degeneracy conditions are exact mathematical condition and are very sensitive to losses. Here we demonstrate an effective approach for observing high *Q*-factors even in the presence of losses.

Against this background, the pervasive concept of exceptional points (EPs) has emerged to describe points of state eigenvector coalescence in coupled circuits, resonators and waveguides with gain and loss. The notion of Parity-Time ($\mathcal{PT}$)-symmetry is widely used to design optical component and lasers [32,33] having spatially-symmetric distributions of gain and loss. These features occur in strictly non-Hermitian systems, which means that the local evolution of waves in the coupled system is described by a non-Hermitian matrix. Therefore, the EP is induced thanks to the presence of gain and/or loss [34–36]. Instead, analogous characteristics of EP could be found in lossless periodic structures at points of degeneracy, e.g., at the SIP, or at the DBE. Nonetheless, the evolution equations of locally-coupled waves in lossless waveguides with DBE or SIP (similar to the ones proposed in this paper) constitute Hermitian matrices. The EP induced in these periodic structures are due to periodicity thanks to which waves can be periodically mixed. From here onward, we denote the EP that is associated to eigenwave coalescence by the acronym EPD to avoid ambiguities since the term "exceptional" may have different meanings in different disciplines, and indeed here we investigate points of degeneracy. Therefore, the EPD is defined here as the point in the parameter space of the CROW at which a degeneracy occurs, i.e., two or more physical wave eigenvectors coalesce.

In this paper, we reveal novel properties of a CROW that is engineered to exhibit EPDs of various orders. Our proposed theory of CROW with modal degeneracies leads to the observation of unprecedented performance in terms of $Q$-factors which are suitable for many applications such as lasers, high sensitive sensors, $Q$-switching devices, to name few. A great advantage of such CROWs is that they can be easily fabricated using optical lithography and possess high $Q$-factors compared to prior implementations of optical waveguides with DBE as those in [27,37,38]. Our paper shows for the first time the DBE condition in CROWs, as well as the SIP. We also demonstrate an anomalous scaling of $Q$-factor and we also investigate this scaling in the presence of losses.

This paper is organized as follows. First, in Sec. II we describe the proposed CROW coupled to the straight waveguide shown in Fig. 1. Then we detail the transfer matrix formalism adopted for the CROW system in Sec. III. In Sec. IV, we analyze the Floquet-Bloch eigenwaves of the proposed CROW and present the mathematical description of all possible EPDS that may arise in such a guiding structure. Then in Sec. V we introduce an analytic expression of the dispersion relation of the proposed CROW unit cell shown in Fig. 2 and the necessary conditions for such unit cell to exhibit a DBE. In Sec. VI we show with examples the dispersion diagram of various EPDs that manifest in the proposed CROW. Finally, in Sec. VII we show the transfer function behavior of such CROW near various EPDs and we briefly describe the unusual properties of the loaded $Q$-factor of CROW cavities with EPDs even in the presence of losses.

## II. GEOMETRY OF PROPOSED CROW

In this paper, we investigate the emergence of high order EPDs in CROWs shown in Fig. 1(a) where the straight waveguide provides an additional coupling mechanism.**Error! Reference source not found.** It consists of a chain of coupled ring resonators, with outer radius $R$, side-coupled to a uniform optical waveguide. We define the field coupling coefficient (also referred to as cross-coupling coefficient [39]) between two contiguous rings or between the waveguide and a ring as $\kappa$ which quantifies how efficiently the field leaks from one to the other (see definitions in [40]). The coupling is assumed to occur at discrete points of closest proximity (denoted hereafter as the *coupling points*), between adjacent rings and between each ring and the straight waveguide as in Fig. 1. For the sake of generality, the field coupling coefficients between the waveguide and the rings may change from pair to pair of adjacent rings as $\kappa_1$ and $\kappa_2$ while the rings coupling coefficients are alternating between $\kappa'_1$ and $\kappa'_2$, respectively (see Fig. 1). Having $\kappa_1 \neq \kappa_2$ and $\kappa'_1 \neq \kappa'_2$ enables the occurrence of the third order degeneracy (the SIP). We also define a corresponding transmission coefficient at the coupling points denoted by $\tau$. In our analysis we assume, for simplicity without loss of generality, that both $\kappa$ and $\tau$ are real positive values, see pages 120-122 in [41] or Ref. [42]. Also, we are assuming lossless coupling at all the coupled sections [15], i.e.

$$\kappa^2 + \tau^2 = 1 \qquad (1)$$

The waveguides and the rings support eigenwaves whose local phase propagation in the positive/negative $z$-direction is represented by $e^{\pm ink_0 z}$ and the time convention $e^{-i\omega t}$ is implicitly assumed. Accordingly, phase propagation (or attenuation) in the waveguide as well as in the rings is modeled by their effective refractive index, namely $n$ which is defined as the ratio between the propagation wavenumber $k$ of the propagating eigenwave normalized by the wavenumber in free space $k_0 = \omega/c$. Moreover, we assume that only a single transverse eigenwave can be excited inside the rings, in each direction, as well as in the waveguide, and we neglect higher order modes interactions. In addition, we assume that ring resonators do not couple waves of different polarizations [39]. Also, we assume that the effective refractive indices $n$ are frequency independent, justified by the narrowband frequency response investigated here.

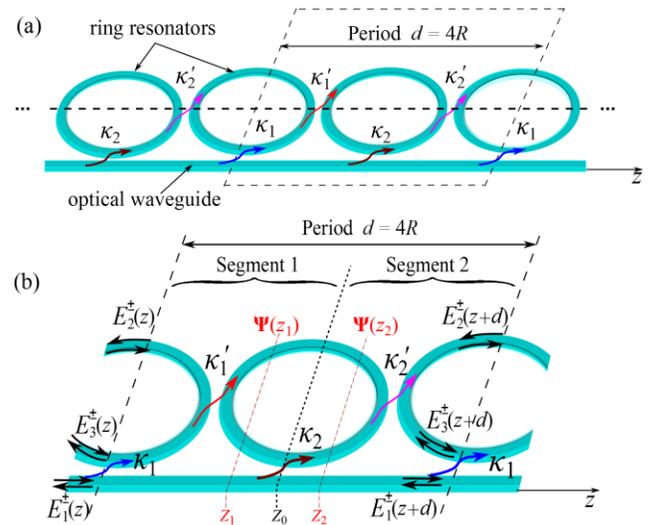

FIG. 1. (a) The proposed CROW is consisting of a chain of coupled ring resonators optical waveguides of radius $R$ side coupled to a rectangular straight waveguide. The field coupling coefficients between the straight waveguide and the rings are alternating between $\kappa_1$ and $\kappa_2$, and the field coupling coefficients between the coupled ring resonators themselves are alternating between $\kappa'_1$ and $\kappa'_2$. The CROW is periodic in the $z$-direction with a period $d = 4R$. (b) the unit cell of this CROW with the electric field wave amplitudes defined at the cell boundaries. $z_0$ is the coupling point between the waveguide and the ring.

The unit cell of the considered periodic CROW is shown in Fig. 1(b) and the CROW's period is $d = 4R$, where $R$ is the outer radius of each ring resonator. We also consider the simplified version with period $d = 2R$ in Fig. 2 that is able to provide the fourth order degeneracy. Note that the coupling between adjacent rings is achieved with the small gaps shown in the figure, however when we evaluate the total length of the period, we neglect such gap lengths as was done in Ref. [15]. As an example, this CROW can be fabricated using silicon on insulator (SOI) technology [43,44]. All geometrical dimensions pertaining to the geometry in Fig. 1 are given in Appendix A; for the cases investigated in this paper.

## III. TRANSFER MATRIX FORMALISIM OF CROWS WITH EXCEPTIONAL POINTS OF DEGENERACY

Analysis of a CROW can be carried out utilizing tight-binding methods [45], transfer matrices [15], or temporal coupled-mode theory [46]; aside from full-wave simulations. In particular, the transfer matrix (T-matrix) method [15,47] is the most convenient and flexible as it naturally allows for the analysis of finite, lossy and dispersive CROWs with strong coupling coefficients that are not necessarily identical for all resonators. Indeed, the transfer matrix is often employed in analysis of layered media [23] as well as investigation of points of degeneracy [21,25]. We adopt the transfer matrix formulation for analyzing the modal behavior as well as the resonance properties of the proposed CROW.

We consider that each waveguide is able to support a single traveling wave along the $+z$-direction and by reciprocity also an analogous travelling wave along the $-z$-direction. These waveguides have uniform cross section and therefore their supported eigenwaves have fixed field distributions. In the straight waveguide, the fields' variation along the $+z$ and $-z$-directions is described by the electric field wave amplitudes $E_1^+(z) \propto e^{+in_w k_0 z}$ and $E_1^-(z) \propto e^{-in_w k_0 z}$, respectively, where $n_w$ is the effective refractive index of the straight waveguide. In the rings, there are four electric field wave amplitudes, $E_2^+(z)$, $E_3^+(z)$, $E_2^-(z)$, and $E_3^-(z)$ as schematically shown in Figs. 1(b) and 2. $E_2^\pm(z)$ are the wave amplitudes in the upper half ring representing waves propagating in opposite directions, and similarly $E_3^\pm(z)$ are the electric field wave amplitudes propagating in the lower half ring (closer to the waveguide).

We then define the electric field wave amplitudes as three-dimensional vectors $\mathrm{E}^+(z) = \begin{bmatrix} E_1^+(z) & E_2^+(z) & E_3^+(z) \end{bmatrix}^T$ and $\mathrm{E}^-(z) = \begin{bmatrix} E_1^-(z) & E_2^-(z) & E_3^-(z) \end{bmatrix}^T$ to represent waves propagating in the $+$ and $-$ $z$-directions, and the superscript $T$ stands for transpose. It is further convenient to represent electric field wave amplitudes at any point $z$ using the six-dimensional *state vector*

$$\mathbf{\Psi}(z) = \begin{pmatrix} \mathrm{E}^+(z) \\ \mathrm{E}^-(z) \end{pmatrix} \qquad (2)$$

which describes how electric waves evolve along the $z$-direction, similar to the formalism in [21,25]. Accordingly, the wave evolution between any two points $z_1$ and $z_2$ (see for example Fig. 1(b)) is governed by the simple translation property

$$\mathbf{\Psi}(z_2) = \underline{\mathrm{T}}(z_2, z_1) \mathbf{\Psi}(z_1) \qquad (3)$$

where $\underline{\mathrm{T}}(z_2, z_1)$ is the 6×6 transfer (T)-matrix that translates the state vector from point $z_1$ to $z_2$. It is interesting to observe the analogy of this formalism with the coupled transmission line approach developed in [48]. This is also a generalization of the coupled-mode formalism developed for conventional CROW [14,15]. It is the subject of the subsequent analysis to obtain $\underline{\mathrm{T}}(z_2, z_1)$ for any $z_2$ and $z_1$.

Note that the structure is reciprocal; therefore, the T-matrix in (3) for any $z_1$ and $z_2$ has a determinant that equals unity, i.e., $\det \left[ \underline{\mathrm{T}}(z_2, z_1) \right] = 1$. In other words the eigenvalues of $\underline{\mathrm{T}}(z_2, z_1)$ must come in reciprocal pairs [21]. Moreover in lossless structures, the T-matrix is J-unitary (as explained in [21,30]), meaning that $\underline{\mathrm{T}}^{-1}(z_2, z_1) = \underline{\mathrm{J}} \, \underline{\mathrm{T}}^\dagger(z_2, z_1) \underline{\mathrm{J}}^{-1}$ where the dagger symbol $\dagger$ denotes the complex-conjugate transpose operation, and the $\underline{\mathrm{J}}$ matrix is given in Appendix B.

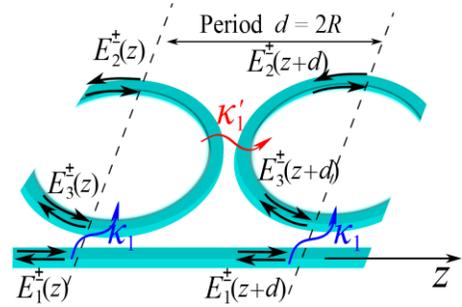

FIG. 2. The CROW unit cell design that supports a DBE and an RBE. The structure is periodic in the $z$-direction with period $d = 2R$. The field coupling coefficient between the waveguide and the resonators chain is $\kappa_1$ while the field coupling coefficient between the coupled ring resonators is $\kappa_1'$. The figure is also showing the orientation of the electric field wave amplitudes at the cell boundaries defined by the dashed lines.

*Scattering matrix (S-matrix) method.* An alternative yet very ubiquitous approach for analyzing CROWs is the use of the scattering matrix or the S-matrix [49]. It is defined by relating the outgoing (or scattered) fields at a specific boundary planes defined at $z_1$ and $z_2$ to the incoming (or incident) fields on the same planes defined at $z_1$ and $z_2$ as in Fig. 1(b). The 6×6 S-matrix is expressed in terms of the electric field wave amplitudes as [49]

$$\begin{pmatrix} \mathrm{E}^-(z_1) \\ \mathrm{E}^+(z_2) \end{pmatrix} = \underline{\mathrm{S}}(z_1, z_2) \begin{pmatrix} \mathrm{E}^+(z_1) \\ \mathrm{E}^-(z_2) \end{pmatrix} \qquad (4)$$

From reciprocity the S-matrix must be symmetric, and for lossless structures (i.e. without dissipative or scattering/radiation losses) the S-matrix must be unitary, i.e. $\underline{\mathbf{S}}\underline{\mathbf{S}}^\dagger = \underline{\mathbf{1}}$ where $\underline{\mathbf{1}}$ is 6×6 identity matrix [50].

We also use the well-known generalized beam splitter relations [51,52] to relate the fields at the coupling points. For example, let us consider the point $z_0$ between a ring and the waveguide as the point where coupling occurs, see Fig. 1(b). Let us define $z_0^-$ and $z_0^+$ to be the $z$-coordinates just before and just after a coupling point at $z_0$, respectively, between the waveguide and the ring resonator. Accordingly, we relate the state vector fields just before the coupling point $E_1^+(z_0^-)$, $E_3^+(z_0^-)$ to those just after the coupling point $E_1^+(z_0^+)$, $E_3^+(z_0^+)$ as

$$\begin{pmatrix} E_1^+(z_0^+) \\ E_3^+(z_0^+) \end{pmatrix} = \begin{pmatrix} \tau_2 & i\kappa_2 \\ i\kappa_2 & \tau_2 \end{pmatrix} \begin{pmatrix} E_1^+(z_0^-) \\ E_3^+(z_0^-) \end{pmatrix} \quad (5)$$

where $\tau_2$ and $\kappa_2$ are the transmission and coupling coefficients, respectively representing the coupling at point $z_0$.

It is intuitive in the proposed CROW to write the wave amplitude evolution equations describing a unit cell using the S-matrix. We proceed by relating the fields at the boundaries of the unit cell, namely at point $z$ and $z+d$, ($d=4R$) as in Fig. 1(b). For simplicity, we divide the unit cell into two segments each of length $2R$. The two segments are similar in construction; the first segment depends on $\kappa_1$ and $\kappa_1'$ whereas the second segment depends on $\kappa_2$ and $\kappa_2'$ as seen in Fig. 1(b). Note that the wave amplitudes at the segment boundaries are defined on the left side of the coupling points; meaning that $\mathbf{\Psi}(z+2R) = \mathbf{\Psi}(z_0^-)$ as can be seen in Fig. 1(b). To determine the scattering matrix of a unit cell, we first calculate the scattering matrix of the individual segments 1 and 2 defined in Fig. 1(b), then we determine their T-matrices and finally we combine them. Consequently, we write the 6×6 S-matrix $\underline{\mathbf{S}}_1$ of the first segment of the unit cell in Fig. 1(b) as

$$\underline{\mathbf{S}}_1 = \begin{pmatrix} \underline{\mathbf{S}}_{11} & \underline{\mathbf{S}}_{12} \\ \underline{\mathbf{S}}_{12}^T & \underline{\mathbf{S}}_{22} \end{pmatrix} \quad (6)$$

where the superscript $T$ denotes the transpose operation, and each of the submatrices has dimensions of 3×3. The submatrices in (6) are expressed as

$$\underline{\mathbf{S}}_{11} = \begin{pmatrix} 0 & i\kappa_1\tau_1'\Omega_r & 0 \\ i\kappa_1\tau_1'\Omega_r & 0 & \tau_1\tau_1'\Omega_r \\ 0 & \tau_1\tau_1'\Omega_r & 0 \end{pmatrix} \quad (7)$$

$$\underline{\mathbf{S}}_{12} = \begin{pmatrix} \tau_1\Omega_w & -\kappa_1\kappa_1'\Omega_r & 0 \\ 0 & 0 & i\kappa_1'\Omega_r \\ i\kappa_1\Omega_w & i\kappa_1'\tau_1\Omega_r & 0 \end{pmatrix} \quad (8)$$

$$\underline{\mathbf{S}}_{22} = \begin{pmatrix} 0 & 0 & 0 \\ 0 & 0 & \tau_1'\Omega_r \\ 0 & \tau_1'\Omega_r & 0 \end{pmatrix} \quad (9)$$

where $\Omega_r = e^{ik_r\pi R}$ is the phase propagation along half of the ring circumference, $\Omega_w = e^{2ik_wR}$ is the phase propagation inside the uniform straight waveguide for a distance $2R$. Here, $k_r$ is the propagation wavenumber in the waveguide making the ring resonators and it is written as $k_r = n_r k_0$ where $n_r$ is the effective refractive index of that waveguide. Similarly, $k_w$ is the propagation wavenumber in the straight waveguide expressed as $k_w = n_w k_0$. In the same manner one can derive the scattering matrix $\underline{\mathbf{S}}_2$ related to segment 2 in Fig. 1(b) by replacing $\kappa_1$ with $\kappa_2$, $\kappa_1'$ with $\kappa_2'$, $\tau_1$ with $\tau_2$, and $\tau_1'$ with $\tau_2'$.

Using the relation between the S-matrix and the T-matrix (Eq. (B2) in Appendix B), we calculate $\underline{\mathbf{T}}_1 \equiv \underline{\mathbf{T}}(z+2R, z)$ and $\underline{\mathbf{T}}_2 \equiv \underline{\mathbf{T}}(z+4R, z+2R)$ as the T-matrices of segments 1 and 2, respectively, from $\underline{\mathbf{S}}_1$ and $\underline{\mathbf{S}}_2$ (the sub matrices of $\underline{\mathbf{T}}_1$ are given in (B3), Appendix B). Finally, we calculate the unit cell T-Matrix, denoted by $\underline{\mathbf{T}}_U$, simply as

$$\underline{\mathbf{T}}_U = \underline{\mathbf{T}}_2 \underline{\mathbf{T}}_1 \quad (10)$$

The state vector evolves across a unit cell as

$$\mathbf{\Psi}(z+d) = \underline{\mathbf{T}}_U \mathbf{\Psi}(z) \quad (11)$$

where $d$ is the period of the CROW. To find periodic solutions of the state vector in the CROW that behave as $e^{ikd}$ where $k$ is the complex Floquet-Bloch wavenumber, the state vector $\mathbf{\Psi}(z)$ must follow the evolution equation

$$\mathbf{\Psi}(z+d) = \zeta \mathbf{\Psi}(z), \qquad \zeta \equiv e^{ikd} \quad (12)$$

Hence, using (11) and (12) we write the eigensystem equation

$$\underline{\mathbf{T}}_U \mathbf{\Psi}(z) = \zeta \mathbf{\Psi}(z) \quad (13)$$

whose eigenvalues $\zeta = \zeta_n \equiv e^{ik_nd}$, with $n = 1, 2, ..., 6$ are evaluated by solving the dispersion equation

$$D(k,\omega) \equiv \det[\underline{\mathbf{T}}_U - \zeta\underline{\mathbf{1}}] = 0 \quad (14)$$

for complex $k$ [21,48]. Solutions of (14) produce eigenvalues $\zeta_n$ associated with regular eigenvectors $\mathbf{\Psi}_n$ that are linearly independent if $\underline{\mathbf{T}}_U$ is diagonalizable. In the proposed CROW, there are six modal (Floquet-Bloch) wavenumbers of the periodic structure, and if $k_1, k_2,$ and $k_3$ are solutions, reciprocity implies that also $-k_1, -k_2,$ and $-k_3$ are solutions (i.e., if $\zeta$ is an eigenvalue of $\underline{\mathbf{T}}_U$ then $\zeta^{-1}$ is another eigenvalue). This property is useful to determine the number of possible degenerate solutions (that have equal wavenumbers). In this paper we find it is convenient to represent the wavenumbers in the fundamental Brillouin zone (BZ) that is defined here within the range from $kd = 0$ to $2\pi$. Accordingly, the center of the BZ is defined at $k = \pi/d$. Because of periodicity solutions associated to wavenumbers $-k_1$, $-k_2$ and $-k_3$ have also Floquet harmonics $-k_1+2\pi/d$, $-k_1+2\pi/d$, and $-k_1+2\pi/d$ in the fundamental BZ.

We construct the diagonal $3\times 3$ matrix $\underline{\underline{\mathbf{k}}}$ of the Floquet-Bloch wavenumbers as

$$\underline{\underline{\mathbf{k}}} = \begin{pmatrix} k_1 & 0 & 0 \\ 0 & k_2 & 0 \\ 0 & 0 & k_3 \end{pmatrix} \quad (15)$$

We also use $\underline{\underline{\mathbf{\Lambda}}}$ as a $6\times 6$ diagonal matrix whose elements are the eigenvalues $e^{ik_n d}$ via

$$\underline{\underline{\mathbf{\Lambda}}} = \begin{pmatrix} e^{i\underline{\underline{\mathbf{k}}}d} & \underline{\underline{\mathbf{0}}} \\ \underline{\underline{\mathbf{0}}} & e^{-i\underline{\underline{\mathbf{k}}}d} \end{pmatrix} \quad (16)$$

where $\underline{\underline{\mathbf{0}}}$ and $\underline{\underline{\mathbf{1}}}$ are the $3\times 3$ zero and identity matrices respectively. Therefore, it follows that the transfer matrix $\underline{\mathbf{T}}_U$, when diagonalizable, is written as

$$\underline{\mathbf{T}}_U = \underline{\mathbf{V}}\,\underline{\underline{\mathbf{\Lambda}}}\,\underline{\mathbf{V}}^{-1} \quad (17)$$

where $\underline{\mathbf{V}}$ is a $6\times 6$ matrix that represents a non-singular similarity transformation that diagonalizes $\underline{\mathbf{T}}_U$, and is computed using the six regular normalized eigenvectors of $\underline{\mathbf{T}}_U$ as $\underline{\mathbf{V}} = [\mathbf{\Psi}_1 \mid \mathbf{\Psi}_2 \mid \mathbf{\Psi}_3 \mid \mathbf{\Psi}_4 \mid \mathbf{\Psi}_5 \mid \mathbf{\Psi}_6]$. This diagonalization and the existence of six independent eigenvectors is only possible if there exists a complete basis of regular eigenvectors of $\underline{\mathbf{T}}_U$. This is not valid anymore at a degenerate condition as described in the following.

## IV. EXCEPTIONAL POINTS OF DEGENERACY IN CROW

The aforementioned analysis is valid unless an EPD emerges. At an EPD, and only there, the unit cell T-matrix $\underline{\mathbf{T}}_U$ is similar to a Jordan Block or a matrix containing Jordan Blocks, analogously to the cases of stratified media and photonic crystals in Refs. [21,53]. At the EPD, the algebraic multiplicity of an eigenvalue of (13) (i.e., the number of identical roots of (14)), is higher than its geometrical multiplicity (the number of independent eigenvectors associated to that eigenvalue).

Note that the evolution equations of the wave amplitudes in the CROW constitutive waveguides can be described with first order differential coupled-wave equations that can be written in a Hermitian form (in the absence of gain and loss), as conventionally done in coupled-mode theory [40]. Therefore, this lossless system can be locally referred to as Hermitian (in the context of coupled-wave propagation [34,40]), even though the T-matrix (transfer matrix) is not Hermitian. Indeed, in our proposed CROW, in absence of losses and gain, a point of degeneracy in the spectrum of $\underline{\mathbf{T}}_U$ emerges only due to periodicity. Accordingly, the eigenvectors and the generalized eigenvectors [21,30] are found by solving

$$\left(\underline{\mathbf{T}}_U - \zeta\underline{\mathbf{1}}\right)^q \mathbf{\Psi}_q(z) = \mathbf{0}, \quad q = 1, 2, \ldots, m \quad (18)$$

where $\mathbf{\Psi}_q$ is the generalized eigenvector and $m$ is the order of degeneracy, i.e., the number of coalescing eigenvectors at the degeneracy point. Note that at least one regular eigenvector is always present in (18). Homogenous solutions of the state vector of (11) shall be represented in terms of eigenvectors and generalized eigenvectors solution of (18). Solutions represented in terms of generalized eigenvectors algebraically diverge along the $z$-direction as $\mathbf{\Psi}(z) \propto z^{q-1}e^{ikz}\mathbf{\Psi}_q(0)$ with $q = 1,2,\cdots,m$ (see Ch. 7 in [54]). Note that $m$ in our CROW can only take the values of 2, 3, 4, and 6 due to symmetry of the Floquet-Bloch wavenumber solutions (reciprocity). Such points of degeneracy occur in the spectrum of $\underline{\mathbf{T}}_U$ by varying system parameters, like frequency for example, or the coupling parameters.

In the following subsections, we will show the mathematical construction of four different types of degeneracies that may emerge in the dispersion diagram of the proposed CROW.

### A. Second order degeneracy: *m=2*

In the given CROW system three scenarios could occur to develop a second order EPD. At the second order EPD, the unit cell T-matrix $\underline{\mathbf{T}}_U$ contains at least one degenerate eigenvalue and eigenvector (i.e. at least two of the six eigenvalue solutions are equal) while the remaining four eigenvalues are in general distinct (unless they also experience another degeneracy). As such, at a given frequency, a CROW can exhibit one, two or even three second order EPDS. The simplest case is when $\underline{\mathbf{T}}_U$ is similar to a matrix that contains one Jordan block, hence it is casted as

$$\underline{\mathbf{T}}_U = \underline{\mathbf{V}} \begin{bmatrix} \underline{\underline{\mathbf{\Lambda}}}_\beta & \underline{\underline{\mathbf{0}}} \\ \underline{\underline{\mathbf{0}}} & \underline{\underline{\mathbf{\Lambda}}}_g \end{bmatrix} \underline{\mathbf{V}}^{-1}, \quad \underline{\underline{\mathbf{\Lambda}}}_g = \begin{pmatrix} \zeta_g & 1 \\ 0 & \zeta_g \end{pmatrix} \quad (19)$$

where $\underline{\mathbf{V}}$, in this case, is a $6\times 6$ matrix composed of five column regular eigenvectors in addition to a generalized eigenvector corresponding to degenerate eigenvalue solutions $\zeta_g$ with a multiplicity of two. Furthermore, $\zeta_g = \exp(ik_g d)$ where $k_g$ is the Floquet-Bloch wavenumber of the second order degeneracy, and $\underline{\underline{\mathbf{\Lambda}}}_g$ is a $2\times 2$ Jordan block shown in (19). $\underline{\underline{\mathbf{\Lambda}}}_\beta$ is diagonal $4\times 4$ matrix with the four remaining eigenvalues. If $\underline{\underline{\mathbf{\Lambda}}}_g$ is the only Jordan block in $\underline{\mathbf{T}}_U$ then $\zeta_g = \pm 1$ meaning that the degeneracy occurs either at the edge ($k_g = 0$) or at the center ($k_g = \pi/d$) of the BZ.

On the other hand, another scenario could manifest when $\underline{\mathbf{T}}_U$ is similar, at a given frequency, to a matrix having two Jordan blocks. This means that there exists two second order EPDs at that frequency, then $\underline{\underline{\mathbf{T}}}_U$ is casted as

$$\underline{\underline{\mathbf{T}}}_U = \underline{\underline{\mathbf{V}}} \begin{bmatrix} \underline{\underline{\mathbf{\Lambda}}}_\gamma & \underline{\underline{\mathbf{0}}} & \underline{\underline{\mathbf{0}}} \\ \underline{\underline{\mathbf{0}}} & \underline{\underline{\mathbf{\Lambda}}}_b & \underline{\underline{\mathbf{0}}} \\ \underline{\underline{\mathbf{0}}} & \underline{\underline{\mathbf{0}}} & \underline{\underline{\mathbf{\Lambda}}}_c \end{bmatrix} \underline{\underline{\mathbf{V}}}^{-1}, \quad \underline{\underline{\mathbf{\Lambda}}}_{b,c} = \begin{pmatrix} \zeta_{b,c} & 1 \\ 0 & \zeta_{b,c} \end{pmatrix} \quad (20)$$

where $\underline{\underline{\mathbf{\Lambda}}}_b$ and $\underline{\underline{\mathbf{\Lambda}}}_c$ are two Jordan blocks corresponding to two pairs of degenerate eigenvalues $\zeta_b$ and $\zeta_c$ respectively, each pair is associated a two second order EPD, one at $k_b$ and the other one at $k_c$. The remaining 2×2 block $\underline{\underline{\mathbf{\Lambda}}}_\gamma$ in (20) is a diagonal matrix that has two eigenvalues that are not related to the degeneracies. Note that if the two 2$^{nd}$ order EPDs occur at the band edge or center, then $\zeta_b$ and $\zeta_c$ must take values of either 1 or −1. Otherwise, the two 2$^{nd}$ order EPDs occur inside the BZ, then $\zeta_b = 1/\zeta_c \neq \pm 1$ (this case will be shown later on in Sec. V, and in Fig. 3).

Furthermore, the last possible scenario is when three second-order degeneracies occur in $\underline{\underline{\mathbf{T}}}_U$, all at a given frequency. Then $\underline{\underline{\mathbf{\Lambda}}}_\gamma$ in (20) is also a 2×2 Jordan block and that could only happen if the eigenvalues $\underline{\underline{\mathbf{\Lambda}}}_\gamma$ are such that $\zeta_\gamma = \pm 1$ i.e., they are either at the BZ edge or center.

### B. Third order degeneracy: *m*=3

When a third order EPD occurs, $\underline{\underline{\mathbf{T}}}_U$ contains two sets of degenerate eigenvalues of algebraic multiplicity 3 and geometrical multiplicity 1, denoted by $\zeta_s$ and $\zeta_s^{-1}$. The T-matrix $\underline{\underline{\mathbf{T}}}_U$ is represented as

$$\underline{\underline{\mathbf{T}}}_U = \underline{\underline{\mathbf{V}}} \begin{bmatrix} \underline{\underline{\mathbf{\Lambda}}}_s^- & 0 \\ 0 & \underline{\underline{\mathbf{\Lambda}}}_s^+ \end{bmatrix} \underline{\underline{\mathbf{V}}}^{-1}, \quad \underline{\underline{\mathbf{\Lambda}}}_s^\pm = \begin{pmatrix} \zeta_s^{\pm 1} & 1 & 0 \\ 0 & \zeta_s^{\pm 1} & 1 \\ 0 & 0 & \zeta_s^{\pm 1} \end{pmatrix} \quad (21)$$

where $\underline{\underline{\mathbf{V}}}$ is composed of two regular eigenvectors and four generalized eigenvectors. The matrices $\underline{\underline{\mathbf{\Lambda}}}_s^+$ and $\underline{\underline{\mathbf{\Lambda}}}_s^-$ are 3×3 Jordan blocks and each degenerate eigenvalue solution, $\zeta_s$ or $\zeta_s^{-1}$, has algebraic multiplicity of 3 and geometrical multiplicity 1. Therefore, there are a regular eigenvector and two generalized eigenvectors associated to each eigenvalue. Here $\zeta_s = \exp(ik_s d)$ and $\zeta_s^{-1} = \exp(-ik_s d)$. Indeed, because of reciprocity that implies symmetries in the dispersion diagram, if $k_s$ is a third-order EPD, then also $-k_s$ must be. In this paper, we show them at $k_s$ and $-k_s + 2\pi/d$. The third order degeneracy is often called SIP resulting in an inflection point in the dispersion curve.

As we discuss in the next sections and in Appendix C, we anticipate that the SIP can be obtained with the unit cell in Fig. 1(b). However, it cannot be obtained using a CROW with a simpler unit cell shown in Fig. 2.

### C. Fourth order degeneracy: *m*=4

When a fourth order EPD occurs, $\underline{\underline{\mathbf{T}}}_U$ contains four degenerate eigenvalues (i.e. four of the six eigenvalue solutions are equal) while the remaining two eigenvalues can be distinct or degenerate. In this case $\underline{\underline{\mathbf{T}}}_U$ is represented as

$$\underline{\underline{\mathbf{T}}}_U = \underline{\underline{\mathbf{V}}}^{-1} \begin{bmatrix} \underline{\underline{\mathbf{\Lambda}}}_\beta & 0 \\ 0 & \underline{\underline{\mathbf{\Lambda}}}_d \end{bmatrix} \underline{\underline{\mathbf{V}}}, \quad \underline{\underline{\mathbf{\Lambda}}}_d = \begin{pmatrix} \zeta_d & 1 & 0 & 0 \\ 0 & \zeta_d & 1 & 0 \\ 0 & 0 & \zeta_d & 1 \\ 0 & 0 & 0 & \zeta_d \end{pmatrix} \quad (22)$$

where $\underline{\underline{\mathbf{V}}}$ is composed of three eigenvectors and three generalized eigenvectors if there exists only a 4$^{th}$ order EPD. The eigenvalue $\zeta_d = \exp(ik_d d)$ has algebraic multiplicity of 4 and geometrical multiplicity of 1. Therefore, there are one regular eigenvector and three generalized eigenvectors associated to $\zeta_d$. In (22), the matrix $\underline{\underline{\mathbf{\Lambda}}}_d$ is a 4×4 Jordan block that contains the four degenerate eigenvalues. Furthermore, $\underline{\underline{\mathbf{\Lambda}}}_\beta$ is a $2\times 2$ diagonal matrix with the two remaining eigenvalues, that are not affected by the fourth order degeneracy, associated to two other eigenvectors. In terms of wavenumbers, and because of reciprocity the fourth order degeneracy has to occur at both $k_d$ and $-k_d$, and in this paper we show what happens at $-k_d$ by looking at the higher Floquet harmonic at $-k_d + 2\pi/d$. Now, the only way to have both wavenumbers $k_d$ and $-k_d$ of multiplicity four (we recall that the total dimensionality of the system is 6), is that $k_d$ must be either $k_d = 0$ or $k_d = \pi/d$, i.e., either at the edge or center of the BZ, respectively, so that $\zeta_d = \exp(ik_d d) = \exp(-ik_d d)$. In other words, this fourth order degeneracy cannot occur at other points of the BZ in a reciprocal system. It is also possible that the block $\underline{\underline{\mathbf{\Lambda}}}_\beta$ experiences another degeneracy of order 2 at the same frequency.

As we show in the next section, we anticipate that the DBE can be obtained with the simpler unit cell in Fig. 2.

### D. Sixth order degeneracy: *m*=6

When a sixth order EPD occurs, $\underline{\underline{\mathbf{T}}}_U$ contains six degenerate eigenvalues, i.e. all the six eigenvalues of (13) are equal, i.e., the algebraic multiplicity is 6 but the geometrical multiplicity is 1. Therefore, the T-matrix $\underline{\underline{\mathbf{T}}}_U$ is represented as

$$\underline{\underline{T}}_U = \underline{\underline{V}}^{-1} \underline{\underline{\Lambda}}_e \underline{\underline{V}}, \quad \underline{\underline{\Lambda}} = \begin{pmatrix} \zeta_e & 1 & 0 & 0 & 0 & 0 \\ 0 & \zeta_e & 1 & 0 & 0 & 0 \\ 0 & 0 & \zeta_e & 1 & 0 & 0 \\ 0 & 0 & 0 & \zeta_e & 1 & 0 \\ 0 & 0 & 0 & 0 & \zeta_e & 1 \\ 0 & 0 & 0 & 0 & 0 & \zeta_e \end{pmatrix} \quad (23)$$

where $\underline{\underline{V}}$ is now composed of a single regular eigenvector and five generalized eigenvectors corresponding to degenerate eigenvalue solutions $\zeta_e = \exp(ik_e d)$ where $k_e$ is the wavenumber at 6th order EPD and $\underline{\underline{\Lambda}}_e$ is 6×6 Jordan block. The sixth order degeneracy has never been investigated in an optical platform and it is the highest order of degeneracy that can be achieved in the configuration shown in Fig. 1, i.e., associated to a system vector with dimension 6. In terms of wavenumbers and because of reciprocity, the system has to experience an EPD of order six also at $-k_e$ (hence $-k_e + 2\pi/d$). It follows that $k_e$ must be either $k_e = 0$ or $k_e = \pi/d$, i.e., either at the edge or center of the BZ, respectively. In other words, this sixth order EPD cannot occur at other points of the BZ in this reciprocal system.

## V. ANALYTIC DIPSERSION RELATION FOR CROWS WITH DBE

In this section, we show that the proposed CROW is able to support EPDs of various orders through proper tuning of the unit cell parameters. Importantly, we derive analytical formulas for the dispersion relation of the CROW as well as the necessary conditions on the CROW parameters to exhibit a DBE (fourth order EPD).

First, let us consider a CROW design where the period is a single ring, i.e., the coupling coefficients from ring to ring are all identical (i.e., $\kappa_1' = \kappa_2'$) as well as the coupling coefficients between rings and the straight waveguide (i.e., $\kappa_1 = \kappa_2$) as shown in the unit cell depicted in Fig. 2. In this case the corresponding CROW period is $d=2R$. This simplified geometry is also able to support the DBE.

The general $k-\omega$ dispersion equation is obtained from (14), for the simplified unit cell in Fig. 2. After some manipulation (14) is casted in the analytic form

$$D(k,\omega) = F(k,\omega)S(k,\omega) - \Gamma(k,\omega) = 0 \quad (24)$$

where $k$ is the Floquet-Bloch wavenumber and $\omega$ is the angular frequency. The three functions in (24) are:

$$F(k,\omega) = \zeta^2 - 2\zeta\cos(2k_w R) + 1$$
$$S(k,\omega) = \zeta^4 + \left[2 + \left(\frac{2\sin(\pi k_r R)}{\kappa_1'}\right)^2\right]\zeta^2 + 1 \quad (25)$$
$$\Gamma(k,\omega) = g\zeta^5 + h\zeta^4 + f\zeta^3 + h\zeta^2 + g\zeta$$

with $\zeta = e^{ikd}$, and

$$g = -2(1-\tau_1)\cos(2k_w R),$$
$$h = \frac{1}{\kappa_1'^2}\left[2(1-\tau_1)\cos(2\pi k_r R) - \kappa_1^2 + 2\kappa_1^2\kappa_1'^2\right] \quad (26)$$
$$f = \frac{-1}{\kappa_1'^2}\left[2\kappa_1^2\cos\left[2R(\pi k_r - k_w)\right] + 4(1-\tau_1)\tau_1'^2\cos(2k_w R)\right]$$

Recall that for a lossless coupling one has $\tau^2 + \kappa^2 = 1$, with $\tau$ and $\kappa$ being real positive numbers, though (24)-(26) are valid also for the more general case where $\tau$ and $\kappa$ are complex. For the trivial case when the coupling coefficient $\kappa_1 = 0$ and $\tau_1 = 1$, then $\Gamma = 0$ in (24,26); and for this trivial case the dispersion equation (24) is casted simply as $D(k,\omega)|_{\kappa_1=0} = F(k,\omega)S(k,\omega) = 0$, which is the dispersion relation of the two decoupled systems: the conventional CROW [15] made of a chain of coupled rings (and not coupled to the straight waveguide) and the isolated straight waveguide. Indeed, the function $F(k,\omega)$ represents the dispersion relation of the straight waveguide alone, i.e., $F(k,\omega) = 0$, that would generate the waveguide wavenumbers $k = \pm k_w \equiv \pm n_w \omega/c$; hence $k_w$ is the wavenumber of the eigenwaves in the isolated straight waveguide and $n_w$ is the associated effective refractive index. Furthermore, the dispersion relation of the conventional CROW [15] is $S(k,\omega) = 0$. Such conventional CROW cannot develop degeneracies with order higher than two. This fact is evident from its dispersion relation as the function $S(k,\omega) = 0$ cannot have roots for the eigenvalue $\zeta$ with multiplicity greater than 2, which is necessary to have a degeneracy higher than a second order. Hence, the only degeneracy condition in the conventional CROW [15] is associated to the eigenvalue equation at the second order EPD angular frequency, called $\omega_g$, as $S(k,\omega_g) = \left(\zeta^2 - \zeta_g^2\right)^2 = 0$, in which $\zeta_g = \pm 1$ are the two eigenvalues with 2nd order degeneracy that must occur either at the BZ edge or center. The reason behind this limitation in a simple conventional CROW is that the chain of coupled resonators does not allow coupling between waves propagating in opposite directions inside each ring. In addition, there is a symmetry between waves propagating in the upper and lower halves of each ring. The occurrence of more general EPDs require coupling of waves traveling in opposite directions that is achieved by s*ymmetry breaking* of the conventional CROW. We define *symmetry* here with respect to a plane perpendicular to the plane that contains the rings, which cuts all the rings of the chain in half as shown with a horizontal dashed line in Fig. 1(a). An effective symmetry breaking is achieved through side coupling to the additional straight waveguide that allows for non-trivial mixing of counter propagating eigenwaves in the CROW as shown next.

Now let us introduce a non-vanishing coupling coefficient $\kappa_1$ that represents the strength of coupling between the chain of resonators and the straight waveguide (Fig. 2). In this case $\Gamma(k,\omega) \neq 0$, and rearranging (24) as $F(k,\omega)S(k,\omega) = \Gamma(k,\omega)$, it is apparent that $\Gamma(k,\omega)$ represents the coupling between the individual dispersion relation of the conventional CROW without waveguide and the straight waveguide. This introduces a structural *symmetry breaking* and leads to coupling of eigenwaves of the CROW, thus higher order EPD can be attained. As we show next, the dispersion equation $D(k,\omega)=0$ in (24) can have solutions representing a fourth order EPD; as well as a second order EPD that is not necessarily at the edge or at the center of the Brillouin zone.

*Necessary conditions for DBE*: The characteristic dispersion equation of the system at the fourth order EPD (i.e., the DBE) frequency can be casted in simple way because the CROW has four degenerate Floquet-Bloch eigenwaves with four coincident eigenvalues, i.e., the characteristic equation must have the following term $(\zeta - \zeta_d)^4$ where $\zeta_d$ is the DBE eigenvalue occurring at the DBE angular frequency $\omega_d$. Moreover, the CROW has also two other eigenwaves that are independent of the four degenerate eigenwaves at the DBE and we assume that their wavenumbers are $k_\beta$ and $-k_\beta$, therefore the characteristic dispersion equation must also have the term $\left(\zeta - e^{-jk_\beta d}\right)\left(\zeta - e^{jk_\beta d}\right)$ which is simplified into $(\zeta^2 - 2\cos(k_\beta d)\zeta + 1)$. Finally, at the frequency at which the forth order EPD occurs, the characteristic dispersion equation that comprises all the eigenvalues of the system *at and only at* the DBE angular frequency $\omega_d$ must take the form

$$D(k,\omega_d) = \left(\zeta^2 - 2\cos(k_\beta d)\zeta + 1\right)\left(\zeta - \zeta_d\right)^4 = 0 \quad (27)$$

The formula (27) and the general dispersion equation (24) are polynomials of order 6 in the variable $\zeta$. By equating the coefficients of these two polynomials, we obtain three equations governing the various CROW parameters that must be satisfied (necessary but not sufficient) in order for a DBE to manifest:

$$\pm 2 + \tau_1 \cos(2k_w R) = \cos(k_\beta d)$$

$$\tau_1 \cos(2\pi k_r R) + 6\kappa_1'^2 + \frac{\kappa_1^2}{2} - \kappa_2^2 \kappa_1'^2 - 1 = \mp 4\kappa_1'^2 \tau_1 \cos(2k_w R)$$

$$\tau_1^2 \cos\left[2(\pi k_r - k_w)R\right] + \cos\left[2(\pi k_r + k_w)R\right] =$$
$$= \pm 8\kappa_1'^2 + \tau_1(4\kappa_1'^2 + 2)\cos(2k_w R) \quad (28)$$

where the $\pm$ sign corresponds to the existence of the DBE either at the center of BZ (i.e. $k_d = \pi/d$ or $\zeta_d = -1$) or at the edge of BZ (i.e. $k_d = 0$ or $\zeta_d = 1$). Even though the values of the lossless CROW parameters (i.e., $\kappa_1, \tau_1, \kappa_2, \tau_2, n_r, n_w$ and $R$) are real and some are bound within certain range, e.g., $0 < \kappa_1, \tau_1, \kappa_2, \tau_2 < 1$, infinitely many points in the parameter space of the CROW whose unit cell is in Fig. 2 can be found where a DBE is manifested. However, the necessary and sufficient condition to develop a DBE is that four eigenvectors of (13) coalesce.

On the contrary, as demonstrated in Appendix C there are no points in the parameter space of the lossless CROW unit cell in Fig. 2 that can lead to an SIP. Therefore, for an SIP to manifest, the more general unit cell of the CROW shown in Fig. 1(b) must be considered.

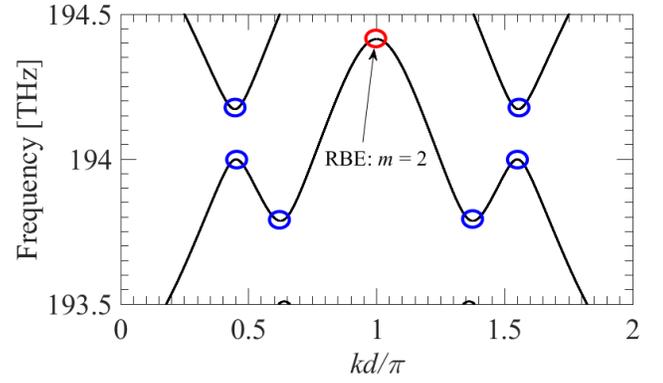

FIG. 3. The Floquet-Bloch wavenumber dispersion diagram associated to a CROW with unit cell shown in Fig.2. It shows various RBEs, one, at the center of the BZ, is at frequency $f \cong 194.41$ THz. The different parameters of the unit cell are set as radius, $R = 10\mu m$, power cross coupling coefficients $\kappa_1^2 = 0.35$, $\kappa_2^2 = 0.469$, and effective refractive indices $n_w = 2.5$, and $n_r = 1.5$.

## VI. EXAMPLES OF DISPERSION RELATION NEAR EPDs IN CROW

In this Section we show how the proposed CROW side coupled to a straight waveguide can exhibit degeneracies of order 2, 4 and 3, called RBE, DBE and SIP, respectively. For the RBE and DBE cases we consider a geometry as in Fig. 2 with parameters given in Appendix A (the same for both cases). For the SIP we consider the geometry in Fig. 1(b) with parameters given in Appendix A.

*Regular Band Edges (RBE).* The CROW constructed from the unit cell depicted in Fig. 2, with period $d = 2R$, develops a second order degeneracy, also called regular band edge (RBE), at the center of the BZ (i.e., $\zeta_g = -1$) marked with a red circle in the CROW Floquet-Bloch wavenumber dispersion diagram depicted in Fig. 3. Note that this plot shows only the propagating eigenwaves of the dispersion diagram in a lossless CROW (i.e. eigenwaves with a purely real Floquet-Bloch wavenumber $k$ as conventionally done in periodic structures [55]). This RBE emerges at frequency $\omega_g/(2\pi) \cong 194.41$ THz, for the parameters in Appendix A, some are summarized in the caption of Fig. 3. The dispersion around the RBE frequency is approximated by $(\omega_g - \omega) \cong h_g(k - k_g)^2$ where the geometry-dependent fitting parameter is $h_g \sim 0.05\omega_g/k_g^2$ for the parameters given in Appendix A. In fact, the dispersion of a conventional CROW [15] with only coupled rings exhibits an RBE. Furthermore, other 2nd order degeneracies (RBEs) not at the edge/center of the BZ are found. At these points the group velocity also vanishes; however, the eigenvalues at such degeneracies are $\zeta_g \neq \pm 1$ as marked with blue circles in Fig. 3. These RBEs not located at the center or edge of a BZ are developed in the proposed CROW in Fig. 2; they cannot occur in a standard CROW (i.e., CROW not side coupled to a waveguide) for the reasons discussed in Secs. IV and V.

*Degenerate Band Edges (DBEs).*

The proposed CROW whose unit cell with period $d = 2R$ is depicted in Fig. 2 exhibits also a fourth order EPD (i.e., the DBE) when coupling parameters are properly designed, as those in Appendix A. A necessary set of equations that govern the different parameters of the CROW unit cell in Fig. 2 are given in (28). The other necessary and sufficient condition to develop a DBE is that four eigenvectors coalesce, meaning that the transfer matrix contains a 4×4 Jordan block as explained in Sec. IV. Such condition is checked numerically once the set of equations (28) are satisfied.

Here the DBE frequency is $\omega_d/(2\pi) \cong 194.83$ THz as shown in Fig. 4 in which the dispersion relation is depicted, for CROW parameters provided in Appendix A, and some are summarized in the caption of Fig. 4. One should notice the flatness at the DBE point, indeed the dispersion around the DBE frequency is approximated by $(\omega_d - \omega) \cong h_d(k - k_d)^4$ where the geometry-dependent fitting parameter is $h_d \sim 0.95\omega_d/k_d^4$ for the given parameters in Appendix A. The CROW with this kind of unit cell can develop either an RBE or DBE at any desired frequency through proper tuning of the different unit cell parameters (mainly the coupling coefficients and the effective refractive indices).

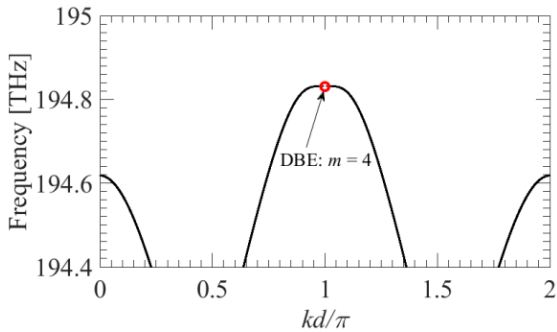

FIG. 4. The Floquet-Bloch wavenumber dispersion diagram of a CROW with unit cell as in Fig.2. It shows a DBE at frequency $f \cong 194.83$ THz and an RBE in the shown frequency range at $f \cong 194.62$ THz. The different parameters of the unit cell are set as in Fig 3.

*Stationary Inflection Points (SIPs).*

As mentioned before, the proposed CROW can develop different kinds of degeneracies. Here, we illustrate how the CROW develops also a third order degeneracy (i.e., an SIP) in its Floquet-Bloch wavenumber dispersion diagram. The unit cell shown in Fig. 2 has been capable to develop RBE and DBE but is not capable of developing an SIP in its dispersion diagram, due to insufficient mixing of waves propagating in the CROW (see Appendix C for a mathematical proof). Therefore, to develop an SIP we use the unit cell with period $d = 4R$ shown in Fig. 1(b) with alternating field coupling parameters among contiguous rings. The parameters of such unit cell are given at the end of Appendix A.

Fig. 5(a) shows the dispersion diagram for this unit cell and two SIPs are obtained at frequency $f \cong 197.04$ THz, symmetrically located with respect to the center of the BZ at $k = \pi/d$. The dispersion around the SIP frequency is approximated by $(\omega - \omega_s) \cong h_s(k - k_s)^3$ where $h_s$ is a geometry-dependent fitting parameter.

Note that Fig. 5(a) shows only branches with purely real wavenumber $k$. However, a more complete picture is provided by plotting the dispersion diagram allowing $k$ to be complex. This is plotted in Fig. 5(b) showing both real and imaginary parts of the Floquet-Bloch wavenumber $k$, where the purely real branches which representing propagating eigenwaves (with zero imaginary parts) are denoted by red lines. From this complex dispersion diagram, one can observe that at each frequency there are six complex values, and how a number $m$ of wavenumbers ($m = 2,3,4$, is the order of EPD) converge to a single one at each EPD. In reading this diagram one should keep in mind that both $k$ and $-k$ (hence $-k + 2\pi/d$) are solutions, because of reciprocity. Furthermore, because of absence of losses, if $k$ is a solution, then $k^*$ (the complex conjugate) is also a solution.

Note that in this case *all* the aforementioned types of EPDs occur in a small frequency band. In other words, RBE, DBE, and SIP manifest at frequencies very close to each other in the same structure. The same closeness of EPDs is observed in Figs. 3 and 4. This is due to the fact that each ring, because of its large size, support several resonances. Using large (i.e., with multiwavelength size) rings is not necessary to develop EPDs of various order, indeed a chain of coupled resonators with smaller dimension than those in this paper (each with a lower $Q$) would also exhibit EPDs. Another example of a simple optical structure supporting DBE and RBE is shown in [56], where two waveguides are periodically coupled.

## VII. GIANT RESONANCE IN CROW WITH EXCEPTIONAL POINTS OF DEGENERACY

In this section we formulate and calculate the quantities relevant to resonators made of a CROW with *a finite number* of coupled rings, side-coupled to a uniform straight waveguide as in Fig. 6. We explore with examples the transmission coefficient (i.e., the transfer function) and the loaded $Q$-factor of the CROW near EPDs of various order as well as the field amplitude distribution inside the CROW at the resonance closest to the EPD frequency. This resonance is generally the sharpest one and for the DBE case we refer to it as the DBE resonance since it is very close to the DBE frequency. We will also investigate the effect of the waveguide and ring losses on the loaded $Q$-factor.

To obtain a solution for the state vector $\mathbf{\Psi}(z)$ at any $z \in [0, L]$, where $L$ is the length of the finite CROW, for a given excitation, we carry out the following steps. We define $\mathbf{\Psi}_0 \equiv \mathbf{\Psi}(0)$ to be the boundary condition at a certain point $z = 0$. Hence the state vector $\mathbf{\Psi}(z)$ at any other point $z$ can be found using $\mathbf{\Psi}(z) = \underline{\mathbf{T}}(z,0)\mathbf{\Psi}_0$, where $\underline{\mathbf{T}}(z,0)$ is the transfer matrix that translates the field from $z$

= 0 to an arbitrary $z$. Let us assume that $\mathbf{\Psi}_0$ defines the electric field wave amplitudes at the left boundary of the first unit cell as seen in Fig. 6. We consider a CROW made of cascaded identical unit cells, as in Fig. 2, with period $d = 2R$, each described by the transfer matrix $\underline{\mathbf{T}}_U$ that translates the state vector across each unit cell as discussed in Sec. III. Accordingly, we relate the state vector at the end of the CROW $\mathbf{\Psi}(L)$, i.e., at the cell boundary of the last unit cell, to $\mathbf{\Psi}_0$ by

$$\mathbf{\Psi}(L) = \underline{\mathbf{T}}_U^N \mathbf{\Psi}_0 \quad (29)$$

where $N$ is the number of unit cells, and the matrix $\underline{\mathbf{T}}_U^N$ is simply calculated as follows. At any frequency except for those at which EPDs occur we bring $\underline{\mathbf{T}}_U$ to a diagonal form as shown in (17).

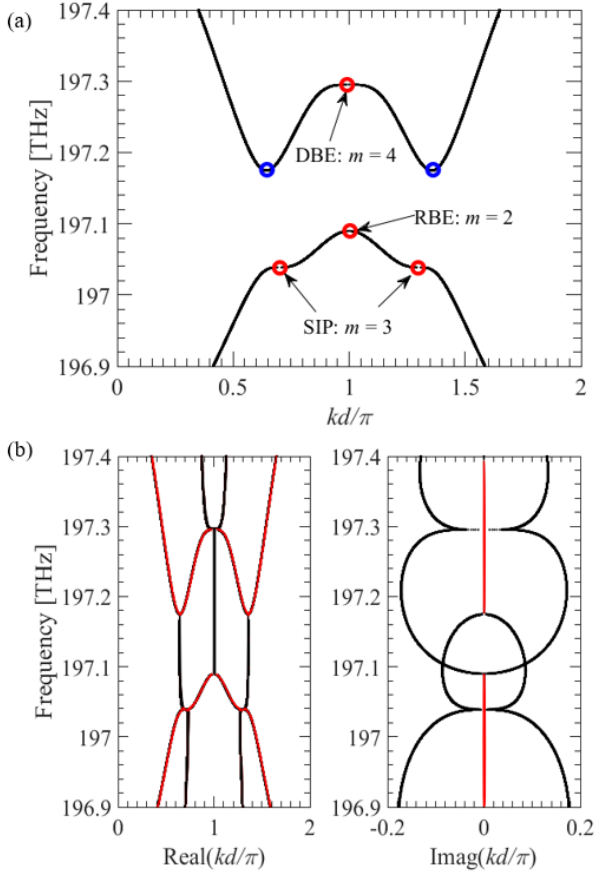

Then matrix multiplication is simply carried out as $\underline{\mathbf{T}}_U^N = \underline{\mathbf{V}} \underline{\mathbf{\Lambda}}^N \underline{\mathbf{V}}^{-1}$. On the other hand, when $\underline{\mathbf{T}}_U$ contains a Jordan Block, i.e., at any of the EPDs discussed in this paper, $\underline{\mathbf{T}}_U$ is non-diagonalizable and based on (20),(21), (22) or (23) one has $\underline{\mathbf{T}}_U^N = \underline{\mathbf{V}} \underline{\mathbf{\Lambda}}_e^N \underline{\mathbf{V}}^{-1}$ where $\underline{\mathbf{\Lambda}}_e^N$ is a matrix containing one or more Jordan Blocks.

*Boundary Conditions.* The state vector at $z = 0$ is chosen to have a specific value $\mathbf{\Psi}(z=0) = \mathbf{\Psi}_0$ dictated by the boundary conditions, i.e., by the load-waveguide attached to the finite CROW and by the left (right) loads attached to the first (last) half-rings as shown in Fig. 6. The six-complex electric field wave amplitudes $\mathbf{E}^\pm(0)$ defined at the boundary of the first ($z = 0$) unit cell and those six $\mathbf{E}^\pm(L)$ at the boundary ($z = L$) of the last unit cell are constrained. We assume that the chain of ring resonators is terminated from both left and right sides with complete rings (recall that the unit cell boundaries choice is made here to start at the middle plane of each ring i.e., at the coupling point, as shown also in Fig. 6). The straight waveguide is extended for $z < 0$ and $z > L$ serving as the feed and the real load of the CROW from which power can be extracted. Therefore, the number of cascaded rings *equals* the number $N$ of unit cells (defined as in Fig. 2) plus one. We assume that the straight waveguide in Fig. 6 is excited by an incoming wave amplitude $E_1^+(0) = E_{\text{inc}}$, coming from the extended waveguide on the left side, i.e., from $z < 0$. Recalling (2), the boundary state vector components at $z = 0$ and $z = L$ are derived as

$$\begin{aligned} E_1^+(0) &= E_{\text{inc}} \\ E_2^+(0) &= e^{ik_2\pi R} E_3^-(0) \\ E_3^+(0) &= e^{ik_2\pi R} E_2^-(0) \end{aligned} \quad (30)$$

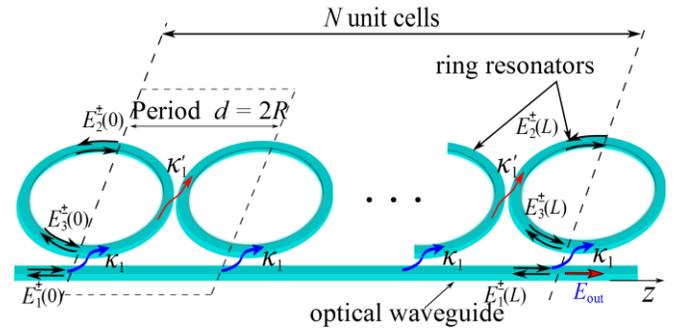

FIG. 6. Finite CROW consisting of $N$ unit cells. We consider the input at $z = 0$ and the output at $z = L$, where $L = 2NR$ and $R$ is the radius of each ring. Note that the large $Q$-factor and the giant scaling with length occur without mirrors, i.e., despite the waveguide coupled to chain of rings is continued to a waveguide with the same dimensions.

and

FIG. 5. (a) The Floquet-Bloch wavenumber dispersion diagram of the CROW unit cell shown in Fig.1(b) showing the three different kinds of EPDs on the same figure. They occur at different frequencies: an RBE at $f \cong 197.1$ THz ; a DBE at $f \cong 197.3$ THz ; and SIP at $f \cong 197.04$ THz . The parameters of the unit cell are chosen as the radius is $R = 10\mu\text{m}$, the power cross coupling coefficients are $\kappa_1^2 = 0.5$ , $\kappa_1'^2 = 0.2$ , $\kappa_2^2 = 0.3$ , $\kappa_2'^2 = 0.4$ , and the effective refractive indices are $n_{\text{eff,w}} = n_{\text{eff,r}} = 1.5$ . (b) The complex dispersion diagram showing both real and imaginary parts of the Floquet-Bloch wavenumber $k$ versus real frequency. Real branches in the complex dispersion (denoted by red lines) represent propagating eigenwaves, while complex branches (denoted by black lines) represent evanescent eigenwaves.

$$E_1^-(L) = i\kappa_1 e^{ik_2\pi R} E_3^+(L)$$
$$E_2^-(L) = t_1 e^{ik_2\pi R} E_3^+(L) \quad (31)$$
$$E_3^-(L) = e^{ik_2\pi R}\left[t_1 E_2^+(L) + i\kappa_1 E_1^+(L)\right]$$

Using the six equations in (30) and (31) along with the six equations of the T-matrix in (29), we solve for the state vector at the boundaries of the finite CROW namely $\Psi_0$ and $\Psi(L)$, and the transfer matrix defined next is also found.

*Transfer function.* We define the transfer function $T_F = E_{out}/E_1^+(0)$ of the proposed CROW as the output electric field wave amplitude $E_{out}$ (see Fig. 6) divided by the CROW excitation $E_1^+(0) = E_{inc}$:

$$T_F = \frac{E_{out}}{E_{inc}}, \quad E_{out} = t_1 E_1^+(L) + i\kappa_1 E_3^+(L) \quad (31)$$

In Fig. 7, we show the transfer function of CROW made of a chain of $N$ unit cells for different values of $N$. The structure is designed with the unit cell in Fig. 2 and parameters in Appendix A to have a DBE, i.e., a fourth order EPD as seen in Fig. 4. We see that for larger number of rings, the transmission peak closest to the 4$^{th}$ order EPD gets narrower (i.e., higher $Q$) and its DBE resonance ($\omega_{r,d}$) gets closer to the DBE frequency $\omega_d$ following the equation $\omega_{r,d} \approx \omega_d - h_d(\pi/Nd)^4$ [21] with $h_d$ being the dispersion fitting parameter and $d$ is the period of the proposed CROW. Furthermore, for a given $N$, the peak closer to $\omega_d$, denoted as resonance $\omega_{r,d}$, is the sharpest one. And this has been used to conceive the single mode of operation of the DBE laser in [56].

In Fig. 8 we show the transfer function of CROW made of a chain of $N$ unit cells for different values of $N$. The structure is designed with the unit cell in Fig. 1(b) with the parameters at the end of Appendix A to have an SIP, i.e., a third order EPD, as seen in Fig. 5. Note that the period is $d = 4R$, therefore the length of the CROW is $L = Nd = 4NR$. We see that the resonance closest to the SIP frequency is getting narrower (higher $Q$) and it is approaching the SIP frequency when $N$ increases. Note that the transmission peak closest to the SIP frequency have a magnitude that is less than unity and this may be attributed to the asymmetric topology of the unit cell, with the respect to the input signal (i.e., the unit cell in Fig. 1(b) is asymmetric about a plane defined at $z_0$ in Fig. 1(b)).

*Quality factor.* We point out that the straight waveguide part of the proposed CROW system *is not* terminated with partially reflecting mirrors at $z = 0$ and $z = L$, and indeed the straight waveguide is just extended without modifications. In other words, the straight waveguide segment part of the CROW is not mismatched when attached to the outside straight waveguide segments, i.e., the same straight waveguide coupled to the chain of rings is continued with the same $k_w$ and $n_w$ for $z < 0$ and $z > L$. Nevertheless, the CROW-waveguide system experiences large mismatch to the outside loading straight waveguides because of the EPD. This is the reason of the high $Q$-factor transmission resonance near the points of degeneracy such as the DBE or SIP shown in Figs. 7 and 8, respectively.

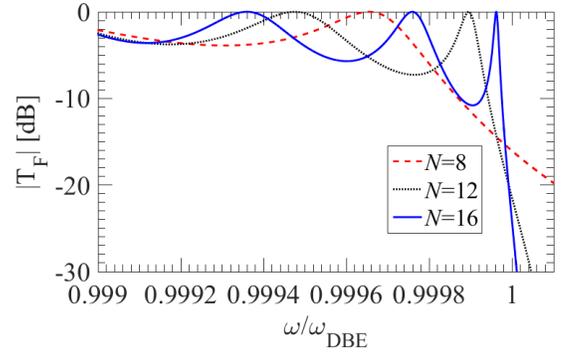

FIG. 7. Magnitude of the transfer function $T_F$ in dB close to a DBE frequency plotted versus angular frequency normalized to the DBE angular frequency ($f_{DBE} \cong 194.83$ THz) for three different number of unit cells ($N = 8, 12$ and $16$) in the CROW with period $d = 2R$.

In general, Floquet-Bloch eigenwaves at frequencies near that of the DBE have characteristic impedance (called the Floquet-Bloch impedance [48]) described by a $3\times3$ matrix that makes the CROW-waveguide system highly mismatched to the terminations impedances (loads) for almost any loading choice. This renders the DBE resonance strongly confined inside the resonator [25,28,57,56], for any load variation. Furthermore, operating near EPDs where the group velocity is vanishing leads to a giant scaling of $Q$-factor with the number of ring resonators in a CROW. The word "giant" has been used to describe this anomalous scaling law encountered also in other geometries supporting the DBE [21,25,28]. This can be inherently understood from the fact that the quality factor is inversely proportional to the group velocity of the Floquet-Bloch wave $v_g$ (i.e. $Qv_g = $ constant) as discussed in [55]. This statement implies that if the DBE resonance $\omega_{r,d}$ coincides with the DBE frequency $\omega_d$ at which $v_g = 0$, then $Q$-factor will be infinite (ideally, in a lossless waveguide). However, the $Q$-factor is finite because $\omega_{r,d} \neq \omega_d$, though they are very close following the equation $\omega_{r,d} \approx \omega_d - h_d(\pi/Nd)^4$. In essence, $(\omega_d - \omega_{r,d}) \propto 1/N^4$, hence the $Q$-factor rapidly increases with the number of cells $N$, because this latter formula describes the rate at which the resonance angular frequency $\omega_{r,d}$ gets closer to $\omega_d$ with growing $N$, which in turn leads to special scaling with $N$ of the increase of group delay and $Q$-factor [58].

The loaded $Q$-factor of the CROW, is estimated numerically using both the fractional bandwidth (FBW) as well as the group delay methods as discussed next. The FBW is defined as the 3dB bandwidth of the transfer function resonance normalized by the resonance frequency, and the $Q$-factor is then calculated as $Q = 1/\text{FBW}$. On the other hand, the group delay is calculated as the derivative of the phase of the transfer function, $\angle T_F$, with respect to the angular frequency $\omega$, i.e., $\tau_g = \partial\angle T_F / \partial\omega$ [25]. The *Q*-factor is then calculated by $Q = \omega_{r,d}\tau_g / 2$ in which $\tau_g$ is the group delay [59].

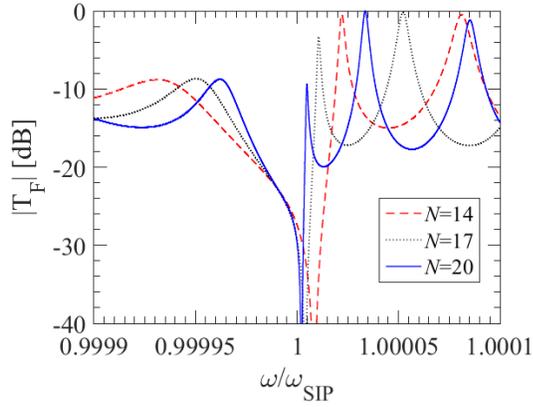

FIG. 8. Magnitude of the transfer function $T_F$ in dB near an SIP frequency plotted versus angular frequency normalized to the SIP angular frequency ( $f_{SIP} \cong 197.04$ THz ) for three different number of unit cells (*N*) in the CROW given as 14, 17 and 20. Note that the unit cell in this case has a period $d=4R$. Hence, the number of rings in this CROW equals $2N+1$.

In Fig. 9, the *Q*-factor of a loaded CROW system, i.e., the CROW side coupled to the straight waveguide continued to the external straight waveguide, is calculated near the DBE frequency, namely at $\omega_{r,d}$, using both methods just discussed, for different number of rings.

The loaded *Q*-factor is normalized to the *Q*-factor of a single ring resonator coupled to the waveguide ($Q_0$) calculated using the well-known equation $Q_0 \approx \omega_{res}\tau_{ph} / (\alpha L_r - 2\ln\tau)$ [60] where $\omega_{res}$ is the ring resonance angular frequency, $L_r = 2\pi R$ is the circumference of ring resonator, $\tau_{ph}$ is the phase delay given by $\tau_{ph} = n_r L_r / c$, and α is the power loss parameter per unit length, that includes radiation and material dissipation. In Fig. 9 we assume that $\alpha = 0$, i.e., rings are lossless. With the parameters in Appendix A, one has $Q_0 \sim 900$. It is clear from Fig. 9 that the two methods used to calculate the loaded *Q*-factor are almost equivalent (for instance see Ref. [59]). The values of normalized *Q* are then fitted to a curve whose equation is $bN^5$ for $N > 9$, resulting in an error described by the parameter $R-\text{squared} \sim 0.995$, where $R$-square is a statistical measure expressing how close the calculated *Q* values are to the fitting curve [61] and it takes values between 0 (worst fitting) and 1 (best fitting). Also in the same figure and using the same unit cell parameters, the normalized *Q*-factor of a loaded CROW working near an RBE frequency is shown and the values are fitted to an equation $bN^3$ with $R-\text{squared} \sim 0.999$. We stress that the *Q*-factor near a DBE scales as $N^5$, which is an unprecedented scaling law with size.

Note that the proposed CROW exhibit both RBEs and DBEs, and we consider the two denoted by the red circles in Fig. 3 and 4. The RBE and DBE are at two distinct, but close, frequencies, 194.22 THz and 194.83 THz respectively. The *Q*-factor evaluated in Fig. 9 is evaluated at the resonance of the finite-length CROW closest to the respective EPD. Fig. 9 also show the scaling of the *Q*-factor associated to the RBE, calculated at its closest resonance, that scales as $N^3$. Both scaling factors, of the DBE and RBE cases, should be compared with the *Q*-factor for a single ring resonator that increases linearly with the length.

*Losses in CROW.* We investigate now the effect of losses on the CROW-waveguide system that includes radiation (due to bending) and dissipative losses. We assume that losses are represented by the attenuation constant of the waveguide and ring resonators. The values of the dissipative losses; as well as radiation losses due to bending, are taken from [62]. Therefore, propagation in the straight waveguide and ring waveguides is characterized by wavenumbers with complex values $k_{r,w} = \text{Re}[k_{r,w}] + i\alpha_{r,w}$, where $\alpha_w = 0.37$ dB/mm is the wave attenuation constant for the straight waveguide whereas $\alpha_r = 0.45$ dB/mm is the attenuation constant inside the rings and it includes dissipative losses 0.37 dB/mm and radiation losses (due to bending) that is assumed to be 0.08 dB/mm (i.e, 0.005 dB/turn for a ring resonator with $R=10\mu$m) [62].

In Fig. 10 we show the effect of these internal losses on the scaling of *Q*-factor for both the RBE and the DBE cases. From Fig. 10 we see that the normalized *Q* is higher for the DBE case. We also observe that the *Q*-factor relative to the DBE case is affected by losses more than the *Q*-factor associated to the RBE case (the RBE and DBE designs are at two different, but close, frequencies, 194.22 THz and 194.83 THz respectively). For CROW-waveguide systems made of a small number of rings the lossless and lossy cases exhibit, more or less, the same *Q*-factor, for both the DBE and RBE cases. The *Q*-factor, in the lossy case, increases till it reaches a saturation value that is attained for the DBE case at a smaller number of rings than for the RBE case. Nevertheless, the *Q*-factor for the DBE case is always higher than that for the RBE case even when considering losses. The scaling of *Q*-factor still exists even in the presence of losses as shown from the microwave experiment [63].

*Resonance with structured field.* The distribution of the electric field wave amplitudes $|E_n(z)| = |E_n^+(z) + E_n^-(z)|$ in each $n=1,2,3$ waveguide paths is shown in Fig. 11, evaluated at discrete *z*-points, one per unit cell, specifically at the left boundary of each unit cell of the proposed CROW in Fig. 2, for both the lossless and lossy cases. For comparison, we show the fields at both the 2nd (RBE) and 4th (DBE) order EPDs. The CROW is made of 16 rings and excited by an incoming guided wave from the left with magnitude $|E_1^+(0)|$, that is used as normalization in Fig. 11. It is important to note that in the DBE case, $|E_n(z)/E_1^+(z)|$ are twice more than the RBE case, consistent with the higher *Q*-factor in Fig. 10. Indeed, based on Fig. 10 it is expected that the DBE case

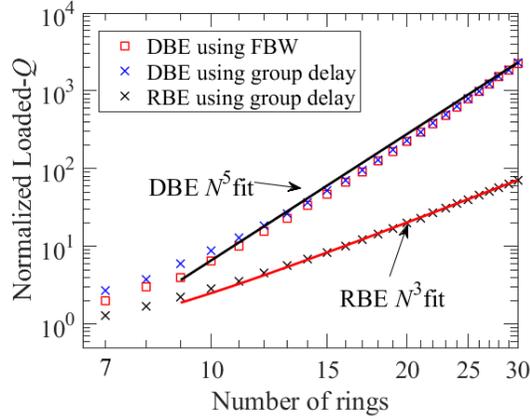

FIG. 9. Loaded quality factor ($Q$) of CROW-waveguide system (without internal losses) normalized with respect to that of a single ring $Q_0$, calculated varying the number of rings for both the RBE and the DBE cases. The values of $Q$, denoted by cross symbols are calculated using the group delay method for DBE and RBE, respectively. However, we also used the fractional bandwidth (FBW) method in the DBE case, represented by square symbols) for the sake of comparing both methods.

Continuous lines represent fitting curves: $N^3$ for the RBE case and $N^5$ for the DBE case, where $N$ is the number of unit cells. Note that the number of CROW rings = $N + 1$.

leads to even stronger field values than the RBE cases for larger number of rings. Finally, we note that the effect of losses in the RBE case is less than that in the DBE case and that is consistent with the aforementioned observations made on the $Q$-factor. Nevertheless, the electric field wave amplitudes are still much stronger in the DBE case even when losses are present.

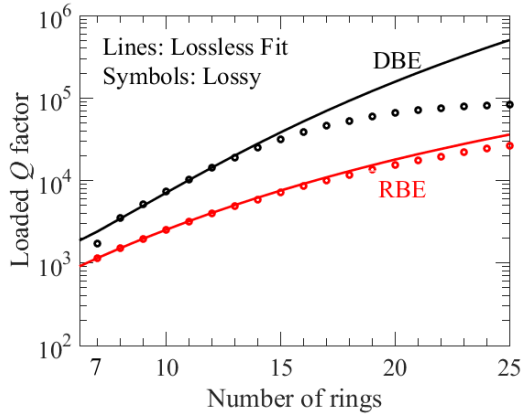

FIG. 10. Calculated loaded quality factor ($Q$) plotted versus number of rings for the lossless and lossy CROW. Each case is evaluated at the resonance frequency closest to the frequency at which the relative EPD occurs (i.e., at $\omega_{r,d}$ for the DBE case). The lines denote the $N^5$ and $N^3$ trends for the lossless case as in Fig. 9; while symbols are the calculated $Q$-factor for lossy CROWs. The number of CROW rings = $N + 1$.

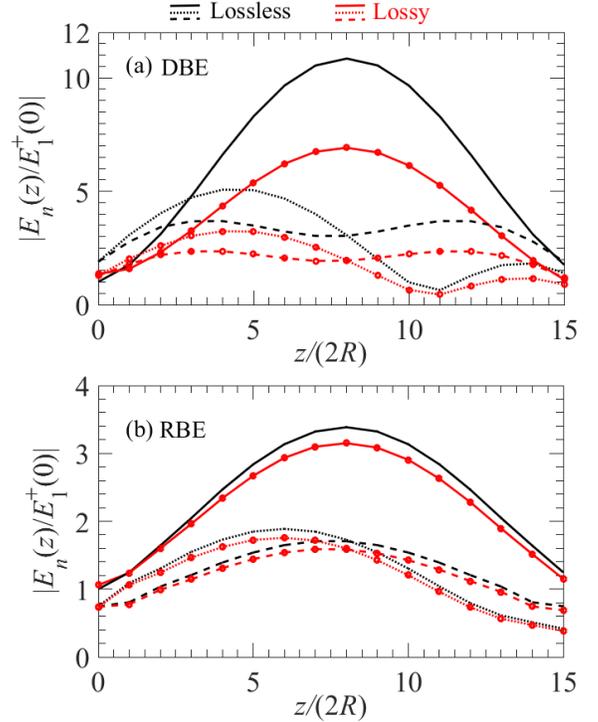

FIG. 11. (a) Absolute values of the total electric field wave amplitude $|E_n(z)| = |E_n^+(z) + E_n^-(z)|$, where $n = 1,2,3$, calculated at one point per unit cell (at the left boundary of each unit cell) in both lossless and lossy CROW-waveguide systems made of 16 rings operating at the DBE resonance $\omega_{r,d}$. (b) As in part (a) but at the RBE resonance. In both cases, solid lines represent field amplitudes in the straight waveguide $|E_1(z)| = |E_1^+(z) + E_1^-(z)|$. Dashed and dotted lines represent the wave amplitudes inside the upper and lower branches of the ring resonators at the left boundary of each unit cell. DBE field values are always higher than RBE electric field wave amplitude values. Moreover, the DBE case with loss has higher field than the RBE case without

## VIII. CONCLUSION

We have presented the concept of a CROW made of a chain of coupled ring resonators, all coupled to an adjacent straight waveguide, that exhibits EPDs with various orders of degeneracy. We have provided the theoretical formulation relying on the transfer matrix analysis that describes such EPDs of various orders. We have demonstrated that EPDs of various orders exist without the need for gain and/or loss. This is in contrast to what has been currently proposed in the context of $\mathcal{PT}$-symmetric optics.

We have also discussed the necessary conditions to realize a DBE in a CROW-waveguide system. We have derived an analytical expression for the CROW dispersion relation, based on the unit cell in Fig. 2, that provides physical insight into the DBE's occurrence. Furthermore, we have shown that the simplest geometry in Fig. 2 is not able to generate SIP degeneracies, that have been obtained instead using the more general unit cell in Fig. 1(b).

Moreover, we have shown that manipulation of various order of degeneracies in such a CROW system leads to unprecedented enhancement in the $Q$-factor and its anomalous scaling law with the CROW length. Such anomalous and unprecedented scaling law has been demonstrated even in the presence of loss. Importantly, when operating at the DBE resonance, the proposed CROW with losses has larger $Q$-factor than the ideal lossless RBE case.

The concepts here discussed are general and the analysis can be applied to several other structures made of multiple coupled waveguides. The examples discussed in this paper involved a CROW that has shown several EPDs at frequencies close to each other because we have used large (in terms of wavelength) resonators that individually support various modes. Another example of coupled waveguides that support DBE and RBE is shown in [56].

Our proposed concept of degeneracies for a CROW coupled to a straight waveguide serves as a promising testbed for enhancing the $Q$ of resonators, even without mirrors, and to have systems whose $Q$ and transfer function changes abruptly by the slight variation of some system parameter, hence making them suitable for extremely sensitive sensors. EPDs like those discussed in this paper are potentially useful toward various applications including non-linear light manipulation and transport, lasers, switches, modulators, and extremely sensitive sensors.

## ACKNOWLEDGMENT

The authors would like to thank Prof. A. Figotin at the University of California Irvine for fruitful discussions. This material is based upon work supported by the Air Force Office of Scientific Research under award number FA9550-15-1-0280 and under the Multidisciplinary University Research Initiative award number FA9550-12-1-0489 administered through the University of New Mexico.

## APPENDIX A: PARAMETERS USED IN NUMERICAL SIMULATIONS

The radius of all the ring resonators considered in this paper is $R = 10 \mu m$, similar to a design demonstrated in [64]. The various parameters of the unit cell shown in Fig. 2 that is able to develop RBE and DBE are as follows: power cross coupling coefficients are $\kappa_1^2 = 0.35$, and $\kappa_1'^2 = 0.469$, while the effective refractive indices that describe propagation in the waveguide and in the rings are set as $n_w = 2.5$ [65], and $n_r = 1.5$. Note that the values of the coupling coefficient were chosen in the range presented in [66] and these values are compatible with SOI fabrication as shown in [67].

For the SIP case, the parameters of the unit cell shown in Fig. 1(b) are as follows: power coupling coefficients are $\kappa_1^2 = 0.5$, $\kappa_1'^2 = 0.2$, $\kappa_2^2 = 0.3$, and $\kappa_2'^2 = 0.4$, while the effective refractive indices are set as $n_w = 1.51$, and $n_r = 1.5$.

## APPENDIX B: TRANSFORMATION FROM S-MATRIX TO T-MATRIX

We show here how to transform the scattering S-matrix into the transfer T-matrix. The advantage of using the T-matrix is that we can characterize our multiple unit cells structure by simply multiplying the T-matrices of the cascaded unit cells [68]. Recall that the transfer matrix relates the state vector at points $z_1$ and $z_2$ as seen from (3). On the other hand, the S-matrix relates them using (4). The following transformation from an S-matrix to a T-matrix is a general property that is applicable to any dimensionality of the system. However, we apply it for any T-matrix or S-matrix relating electric field wave amplitudes at an arbitrary points $z_1$ and $z_2$ in the proposed CROW. Therefore, the T-matrix and S-matrix, for our proposed CROW, are $6 \times 6$ matrices. We represent the $6 \times 6$ S-matrix and T-matrix using four sub-block matrices, each of size $3 \times 3$, as

$$\underline{S}(z_2, z_1) = \begin{pmatrix} \underline{\underline{S}}_{11} & \underline{\underline{S}}_{12} \\ \underline{\underline{S}}_{21} & \underline{\underline{S}}_{22} \end{pmatrix}, \quad \underline{T}(z_2, z_1) = \begin{pmatrix} \underline{\underline{T}}_{11} & \underline{\underline{T}}_{12} \\ \underline{\underline{T}}_{21} & \underline{\underline{T}}_{22} \end{pmatrix} \quad (B1)$$

Now we transform the sub-blocks of the S-matrix into the T-matrix sub-blocks through the following expressions [68]

$$\begin{aligned} \underline{\underline{T}}_{11} &= \underline{\underline{S}}_{21} - \underline{\underline{S}}_{22} \underline{\underline{S}}_{12}^{-1} \underline{\underline{S}}_{11} \\ \underline{\underline{T}}_{21} &= -\underline{\underline{S}}_{12}^{-1} \underline{\underline{S}}_{11}, \quad \underline{\underline{T}}_{12} = \underline{\underline{S}}_{22} \underline{\underline{S}}_{12}^{-1} \\ \underline{\underline{T}}_{22} &= \underline{\underline{S}}_{12}^{-1} \end{aligned} \quad (B2)$$

Accordingly, after proper substitutions of the S-matrix sub-blocks given in (B1) into (B2), the transfer matrix of the first segment $\underline{T}_1$ of the unit cell shown in Fig. 1(b) is given by its sub-block matrices viz

$$\underline{\underline{T}}_{11} = \underline{\underline{T}}_{22}^* = \begin{pmatrix} \tau_1 \Omega_w & 0 & i\kappa_1 \Omega_w \\ \dfrac{\kappa_1}{\kappa_1'} \Omega_r & 0 & i\dfrac{\tau_1}{\kappa_1'} \Omega_r \\ 0 & i\dfrac{\Omega_r}{\kappa_1'} & 0 \end{pmatrix},$$

$$\underline{\underline{T}}_{12} = \underline{\underline{T}}_{21}^* = \begin{pmatrix} 0 & 0 & 0 \\ 0 & -i\dfrac{\tau'}{\kappa_1'} & 0 \\ -\dfrac{\kappa_1 \tau_1'}{\kappa_1'} & & -i\dfrac{\tau_1 \tau_1'}{\kappa_1'} \end{pmatrix} \quad (B3)$$

where all these quantities were defined before in Sec. III. Analogous steps are used to derive the transfer matrix of the other segment, shown in Fig. 1(b), through replacing the coupling coefficients $\kappa_1, \kappa_1'$ with $\kappa_2, \kappa_2'$ and the transmission coefficients $\tau_1, \tau_1'$ with $\tau_2, \tau_2'$.

Recall that the lossless condition of the S-matrix is $\underline{S}(z_2, z_1)\underline{S}^\dagger(z_2, z_1) = \underline{1}$, where the dagger $^\dagger$ implies complex conjugation and transpose operation, meaning that it is unitary. In addition, the T-matrix of the lossless CROW obeys the fundamental J-unitary property (similar to general stratified media in [21]); which means that $\underline{\mathbf{T}}^{-1}(z_2,z_1) = \underline{\mathbf{J}}\,\underline{\mathbf{T}}^{\dagger}(z_2,z_1)\,\underline{\mathbf{J}}^{-1}$ with the matrix $\underline{\mathbf{J}}$ here is given by

$$\underline{\mathbf{J}} = \begin{pmatrix} 1 & 0 & 0 & 0 & 0 & 0 \\ 0 & 1 & 0 & 0 & 0 & 0 \\ 0 & 0 & 1 & 0 & 0 & 0 \\ 0 & 0 & 0 & -1 & 0 & 0 \\ 0 & 0 & 0 & 0 & -1 & 0 \\ 0 & 0 & 0 & 0 & 0 & -1 \end{pmatrix} \quad (B4)$$

### APPENDIX C: ON THE NECESSARY CONDITIONS GOVERNING THE CROW PARAMETERS TO DEVELOP AN SIP

The dispersion equation of the proposed CROW, made of periodic rings side coupled to the straight waveguide, whose unit cell is shown in Fig. 2 with period $d = 2R$, is obtained by solving the general expression (14); in which the unit cell transfer matrix $\underline{\mathbf{T}}_U$ has sub-blocks defined in (B3) and it is casted analytically in (24). In the following we consider the lossless CROW parameters ($\kappa_1, \tau_1, \kappa_2, \tau_2, n_w, n_r$, and $R$) that are real, positive and some have restrictions on their values i.e., $0 < \kappa_1, \tau_1, \kappa_2, \tau_2 < 1$.

The necessary conditions governing the various parameters of the lossless CROW unit cell shown in Fig. 2 in order to develop a DBE is given in (28). Contrarily, no points in the parameter space of the lossless CROW unit cell in Fig. 2 can be found such that SIP is observed, and the reason for this statement is laid out in the following. The necessary condition to achieve an SIP is that the characteristic equation of the CROW unit cell in Fig. 2 at an SIP angular frequency $\omega_s$ must take the form

$$D(k,\omega_s) = \left[\zeta - \zeta_s\right]^3 \left[\zeta - \frac{1}{\zeta_s}\right]^3 = 0 \quad (C1)$$

which means that there are two eigenvalues each with multiplicity of order three and they are a reciprocal pair, where the eigenvalue $\zeta_s = e^{ik_s d}$ and $d=2R$. Similar to what we did for the DBE case, by equating the coefficients of this polynomial to those of (24), we get three equations governing the CROW parameters that are necessary to obtain the SIP for the geometry in Fig. 2 as

$$\tau_1 \cos(2k_w R) = 3\cos(2k_s R)$$
$$\cos(2\pi k_r R) = \frac{1}{\tau_1}\left[6\kappa_1'^2 \cos^2(2k_s R) + (1-\tau_1^2)(\kappa_1'^2 - 0.5) + 1\right]$$
$$\tau_1^2 \cos(2R(\pi k_r - k_w)) - 2\tau_1\tau_1'^2 \cos(2k_w R) + \cos(2R(\pi k_r + k_w))$$
$$= -\kappa_1'^2 \left(\cos(6k_s R) + 9\cos(2k_s R)\right) \quad (C2)$$

Omitting the details, the equalities in (C2) that represent the necessary conditions to achieve an SIP for the CROW unit cell in Fig. 2 cannot be simultaneously satisfied for any of the lossless CROW parameters except for the trivial case when $\kappa_1 = 0$. Therefore, we find an SIP using a CROW with the more general unit cell with period $d = 4R$ as in Fig. 1(b) that has more degrees of freedom than that shown in Fig. 2. The necessary condition for such unit cell to develop an SIP could also be derived analogously to what has been shown for the DBE.